\documentclass[5p,times]{elsarticle}

\journal{Optics Communications}
\usepackage[utf8]{inputenc}

\usepackage{graphicx}
\usepackage{xcolor}
\usepackage{subcaption}

\usepackage{amsmath}
\usepackage{amsfonts}
\usepackage{comment}
\usepackage{upgreek}

\usepackage{xcolor}

\makeatletter
\providecommand{\doi}[1]{%
  \begingroup
    \let\bibinfo\@secondoftwo
    \urlstyle{rm}%
    \href{http://dx.doi.org/#1}{%
      doi:\discretionary{}{}{}%
      \nolinkurl{#1}%
    }%
  \endgroup
}
\makeatother
\makeatletter
\def\@author#1{\g@addto@macro\elsauthors{\normalsize%
    \def\baselinestretch{1}%
    \upshape\authorsep#1\unskip\textsuperscript{%
      \ifx\@fnmark\@empty\else\unskip\sep\@fnmark\let\sep=,\fi
      \ifx\@corref\@empty\else\unskip\sep\@corref\let\sep=,\fi
      }%
    \def\authorsep{\unskip,\space}%
    \global\let\@fnmark\@empty
    \global\let\@corref\@empty  
    \global\let\sep\@empty}%
    \@eadauthor={#1}
}
\makeatother

\begin{document}
\title{Invariance property in scattering media and absorption}

\begin{abstract}
\end{abstract}
\begin{frontmatter}

%
\author{Federico Tommasi\corref{cor1}}\cortext[cor1]{Corresponding author}
\ead{federico.tommasi@unifi.it}\author{Lorenzo Fini}\author{Fabrizio Martelli}\author{Stefano Cavalieri}
\address{Dipartimento di Fisica e Astronomia, via Giovanni Sansone 1, I-50019, Sesto Fiorentino, Italy}
%
%

%

\begin{abstract}
In this paper we deal with the influence on absorption of the diffusive media characteristics framing the problem in connection with the invariance property (IP) of the mean path length. We show that the IP is an important issue that regulates but not prevent the search of absorption maximization by scattering characteristics. We find that the scattering may increase the absorption or even be detrimental, depending on the geometry of the medium
and the conditions of its illumination.

\end{abstract}


\end{frontmatter}


The study about light scattering properties of different materials has been extensively carried out and has lead to different applications encompassing different fields of optics, such as tissue optics \cite{0034-4885-73-7-076701,Leff2008,0031-9155-58-11-R37,Zaccanti:03}, random lasers \cite{rl4,rl6,rl7,nostro2,PhysRevA.98.053816}, optical sensing based on scattering \cite{cancer,sensore,Tommasi:18,sens_scat} and imaging \cite{Yilmaz:15}, and also leading to more exotic phenomena \cite{dis_phot}, such as replica symmetry breaking behavior \cite{rsb1,rsb_nuovo,rsb3,rsb4}, anomalous diffusion \cite{lf1,genLB2,PhysRevA.99.063836} and Anderson localization \cite{and_loc}.  A very important property about light propagation through scattering media is the invariance property (IP) of the mean path length,  independently presented by Blanco and Fournier  \cite{blanco} and by Bardsley  and Dubi \cite{dubi}, that is a generalization of the mean chord theorem, also used by Dirac in the field of nuclear physics \cite{Dirac,case_zweifel}. They showed that for any purely diffusing system, under isotropic uniform incidence, the average length of trajectories through the system depends only on the geometry of the system and is independent of the characteristics of the diffusion process. This counterintuitive result states that the mean length $\langle L\rangle$ of the corresponding random walk trajectories  inside a flat  domain, of area $S$ and perimeter $C$, is constant: $\langle L\rangle_{IP}=\pi\,S/P$. A similar relation also holds in the three dimensional case, $\langle L\rangle_{IP}=4\,V/S$,  V being the volume and S is the external surface of the domain. 
The result was triggered by the study of random walks of biological species such as ants but has great consequences in many fields. The surprising element of this result is evident when applied to physical sciences and, in particular, to the transport of light or of other types of waves in scattering media. 
In that context, it is well known that all the relevant observable quantities depend on the  reduced scattering coefficient $\mu_s'$ (the reciprocal of the transport mean free path $\ell_t$), that is connected to the scattering  coefficient $\mu_s$ (the reciprocal of the scattering mean free path $\ell_s$) by the relation  $\mu_s'=\mu_s\,(1-g)$, where $g$ if the asymmetric factor of the scattering phase function.  In the diffusive regime, for instance, the total transmission of a slab of thickness $d$ scales with $\mu_s'\,d$ through the Ohm's law \cite{Sheng}. Recently, it has been theoretically shown that the IP also holds for the scattering in resonant structures as well as in ballistic propagation, or even in condition of 2D Anderson localizzation \cite{Pierrat17765}. The experimental evidence of the IP against a two order of magnitude variation of  $\mu_s'$ has been also achieved in a brilliant experiment on  multiple scattering of light in colloidal suspensions of particles in water \cite{Savo765}.

The  maximization of light absorption in the context of the research on new photovoltaic technologies has been an important field of research. In general, enhanced absorption is provided by engineering the surface and/or the volume of the absorbing medium to increase the path of the light. In this context, it has been studied the use of scattering properties to enhance the interaction of light with the medium and then its absorption \cite{Atwater,Pillai,Mallick,Mupparapu:15}.
These studies are peculiar when the absorption is intrinsically low and then attempts to improve it are crucial for practical applications. Indeed, enhancement of absorption has been found in diffusive materials as in random nanophotonic structures \cite{holes,Pratesi:13}. Recently, also a theoretical study on the enhancement of absorption in stealth hyperuniform disordered media has been presented \cite{Bigourdan:19}.

In this paper we deal with the influence on absorption of the diffusive media characteristics. We will show that the IP is an important issue that regulates but not prevent the search of absorption maximization by scattering characteristics. 
 
We will see that the scattering may increase the absorption or even can be detrimental, depending on the geometry of the medium and the conditions of its illumination. In the following, we first analyse the case of spherical geometry subjected to an isotropic illumination, then we report on the case of the slab with and without reflecting faces in the presence of isotropic or mono-directional illumination. The mono-directional illumination is assumed perpendicular to the surface of the medium.

In this work the scattering is low enough ( $\mu_s\,\lambda\ll 1$) to put the system under study  far away from the condition of localization ( $\mu_s\,\lambda\simeq 1$ ). Hence, here we follow an approach that neglects the effect of interference. 

\section{Analytic approach}
In a basic approach, the absorption along a random path in a domain can be found by using the Lambert-Beer (LB) law with absorption coefficient $\mu_a$. Considering the  possible total path $L$ of the radiation inside the medium as a random variable with probability density function (pdf) $P(L)$, mean $\langle L\rangle$ and variance $\mathrm{Var}(L)$, the total absorption $A$ can be approximated in the limit of low absorption ($\mu_a\langle L\rangle\ll 1$) by (see \ref{appen}): 
\begin{equation}
A\approx \mu_a\,\langle L\rangle-\frac{{\mu_a}^2}{2}\,\left( \langle L \rangle^2 + \mathrm{Var}(L) \right)\,.
\label{formula}
\end{equation}
In the case of isotropic illumination, because of the invariant property above described, within the first order approximation in Eq.~(\ref{formula}), the absorption is independent of the scattering strength. The influence of scattering is then linked to the details of the $P(L)$ and, in the second order approximation, to the variance of the distribution. Here we show that the condition of illumination is crucial to determine the effect of scattering on absorption.

In the case of isotropic (lambertian) illumination, $\langle L \rangle$ is an invariant quantity of the medium independent of scattering. Hence, as indicated by Eq.~(\ref{formula}), a crucial point is played by the behaviour of the variance of the path distribution. If the variance increases with scattering, the absorption of the radiation inside the medium decreases. The absorption is then expected to decrease  when scattering is strong. In fact, in such a case there is a larger probability to have both long and very short paths and this causes the variance to increase. Photons can escape in a very short time near the surface but will stay for a long time inside once they reach the inner part of the volume.

\begin{figure}
\centering\includegraphics[width=\columnwidth]{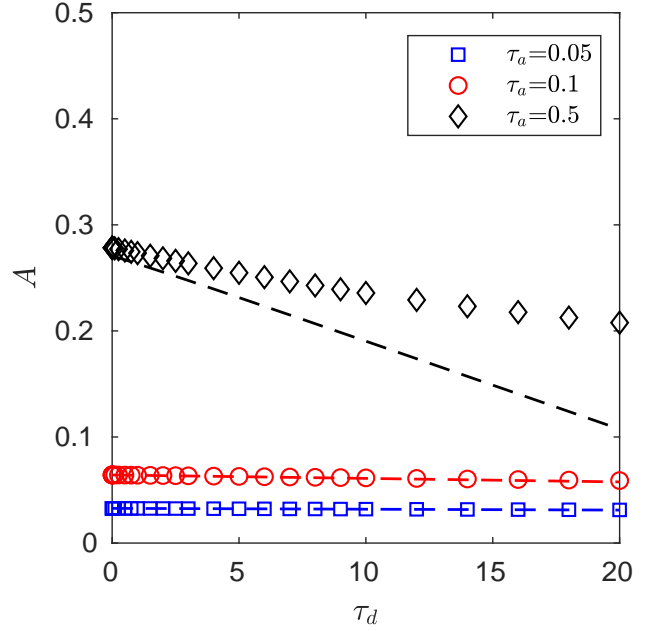}
        \caption{Absorption within a sphere, for different values of $\tau_a$ and in case of lambertian illumination without refractive index mismatch.  The markers pertain to the numerical simulation, while the dashed lines pertain to the corresponding analytical solution  (Eq.~(\ref{formula})).}
        \label{fig:noR_lamb}
   \end{figure}
 \begin{figure}
        \centering\includegraphics[width=\columnwidth]{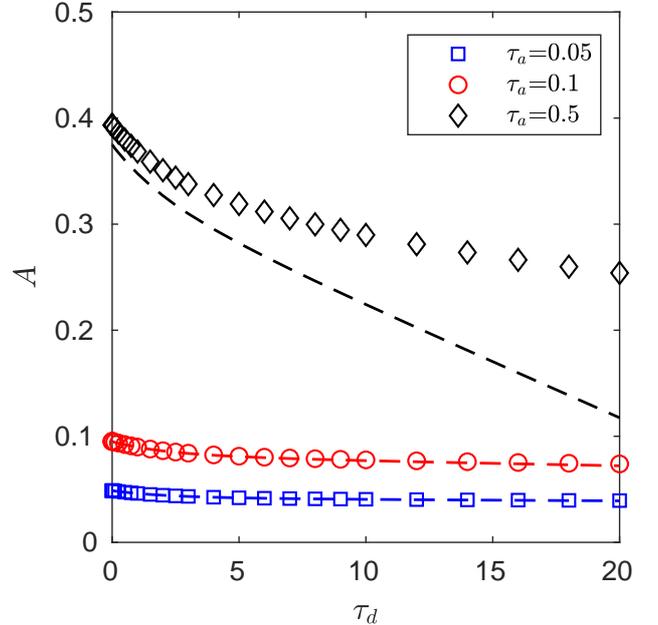}
        \caption{Absorption within a sphere, for different values of $\tau_a$ and in case of mono-directional illumination without refractive index mismatch. The markers indicate  the numerical simulation, while   the dashed lines are the corresponding analytical solution (Eq.~(\ref{formula})).}
        \label{fig:noR_beam}
   \end{figure}
   
Inside a medium of width $d$, the quantity  $\tau_d=\mu_s'\,d$ is  the optical density of the material only due to the scattering, while $\tau_a=\mu_a\,d$ is only due to the absorption.   It is  interesting to note that for strong scattering ($\tau_d\gg 1$) isotropic or mono-directional illumination are expected to  have similar effects; in fact, in that case  the radiation is randomized in a short space and then the resulting behaviour is similar for two illumination conditions. The behaviour is  more unpredictable for low scattering strength. Since in such a case the radiation is not completely randomized,  both the geometry of the medium and the direction of illumination are expected to play a crucial role and then detailed analysis is needed.

 Here we provide more insight of that by considering both isotropic and mono-directional illumination. In particular, in the latter the hypotheses of the IP are not satisfied. The two geometries under  study are the sphere and the infinite laterally extended slab. While in the case of the sphere one has the maximal symmetry and no difference in the three dimensions are present, in the case of the slab the transverse dimensions are supposed much larger than the width. The length $d$ is the thickness in the case of the slab, while it is the diameter in the case of the sphere. The case of refractive index mismatch between the medium and the external environment is also considered.

\section{Numerical simulations}

The numerical simulations are based on a previously developed Monte Carlo (MC) code \cite{Zaccanti:91,Contini:97,Contini:97_2,libro,Sassaroli:98}. The generation of trajectories is the core of the MC program and it is checked to be in excellent agreement with exact analytical expressions \cite{Zaccantietal94}. Each step of length $\ell$ of a trajectory is randomly drawn by the LB probability density function:
\begin{equation}
p\left(\ell\right)=\frac{1}{\ell_s}\exp{[-\mu_s\,\ell]}.
\label{LB}
\end{equation}
Being $\xi\in[0,1]$ a uniformly distributed random number, each step is obtained by the usual inversion of the cumulative distribution associated to the Eq.~(\ref{LB}):
\begin{equation}
\ell\left(\xi\right)=-\frac{\ln{\left[1-\xi\right]}}{\mu_s}
\label{LB2}
\end{equation}
 Each simulation consists in $10^8$ simulated trajectories.
 The Henyey-Greenstein scattering  function is used to generate the scattering angles  
 \cite{hg}.
The results reported in this manuscript are in particular devoted to the study of the case $g=0$, since the main parameter to describe the scattering is $\mu_s'$. Hence, the results for $g\neq 0$ are qualitatively very similar (see Sec.~\ref{secslab}).

\subsection{Spherical geometry}

We start our analysis with a spherical geometry. Absorption versus scattering is reported for isotropic illumination  in Fig.\ \ref{fig:noR_lamb} and mono-directional illumination in Fig.\ \ref{fig:noR_beam}. The results confirm the qualitative considerations reported above for high values of $\tau_d$. A monotonic decreasing of absorption with scattering is evident both for isotropic or mono-directional illumination and for any value of $\mu_a$.  In the figure the analytical expression given by Eq.~(\ref{formula}) is also reported, showing that it is a good approximation for small absorption ($\tau_a<1$). $\langle L\rangle$ and $Var(L)$  can be numerically evaluated by the empirical pdf $\tilde{P}(L)$ created with the data generated by the MC simulations. In this geometry of maximal homogeneity, an increase of the variance leads to the decreasing of absorption also for low values of $\tau_d$. The results indicate then that scattering is detrimental for absorption in case of homogeneous geometry.  

In Fig.~\ref{fig:noR_lamb}, in the case of low absorption ($\tau_a$ = 0.05 and 0.1) and lambertian illumination, the behaviour appears in good approximation
as independent of the scattering. In these cases the
system approaches the sufficient conditions of validity of IP,
that are  uniform and isotropic illumination and the no absorption.
In this conditions the first term in Eq.~\ref{formula} gives the main contribution and it is independent of the scattering.  

\subsection{Slab geometry}\label{secslab}
\begin{figure}
  \centering\includegraphics[width=\columnwidth]{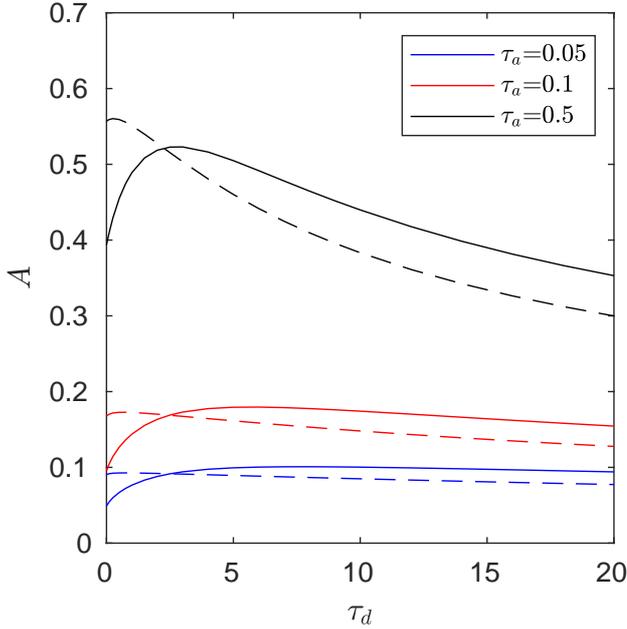}
        \caption{Absorption as a function of $\tau_d$ in the case of slab without refractive index mismatch for different values of  $\tau_a$. The continuous line is with  mono-directional illumination, while the dashed one is with lambertian illumination.}
        \label{fig:slab_noR}
   \end{figure}

In this case the results for the slab geometry are reported on Fig.~\ref{fig:slab_noR}. The absorption is again decreasing for large $\tau_d$ as indicated in the qualitative considerations reported above on the basis of Eq.~\ref{formula}. The behaviour for small $\tau_d$ is more articulate. It is shown a very small increase for isotropic illumination  and a more pronounced maximum for mono-directional illumination ( that is assumed  perpendicular to the surface also for the slab). As $\tau_a\to 0$, in case of lambertian illumination the absorption becomes independent of scattering, as expected when the IP is valid.  

\begin{figure}
        \centering\includegraphics[width=\columnwidth]{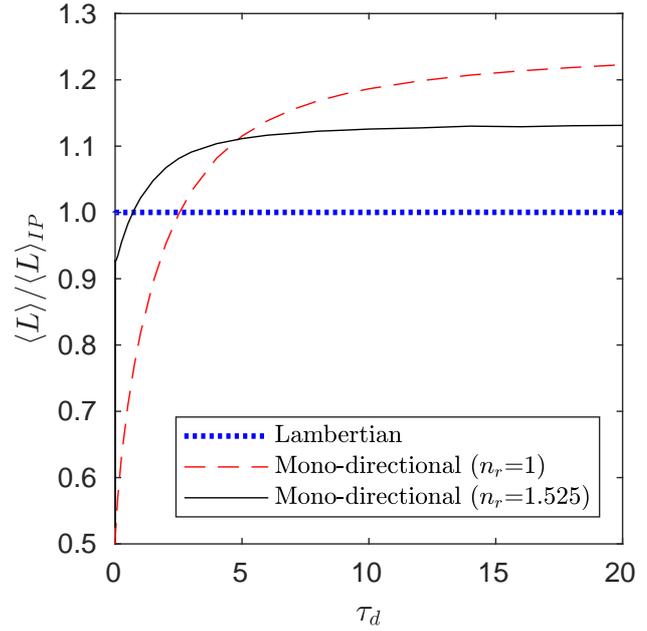}
        \caption{Mean path length normalized to the corresponding invariance property value $\langle L\rangle_{IP}$ (20 mm for $n_r=1$ and 46.5 mm for $n_r=1.525$ in a slab geometry. }
        \label{fig:Lmedio}
   \end{figure}

The last case can be explained by thinking to the role of scattering to have long paths and then greater absorption. The geometry is crucial for this result, since the slab, unlike the sphere, is a geometry with strong differences in extension of the three dimensions.  It is then possible for a scattering event to lead the light from orthogonal to the surface propagation to a much longer path parallel to the entrance surface. This is indeed the case of mono-directional illumination where the invariance property is not more fulfilled. In Fig.\ \ref{fig:Lmedio} is reported the effect of scattering on $\langle L \rangle$ for lambertian and mono-directional illumination. The scattering influence on absorption for mono-directional illumination results then into the balance between increasing due to longer mean path (first order in Eq.~(\ref{formula})) and decreasing due to larger variance (second order in Eq.~(\ref{formula})). In such a plot we also anticipate simulation results in the case of slab geometry with refractive index mismatch.

Many attempts of maximizing absorption have indeed been devoted to deviate light entering a material in long paths inside the slab with propagation near to parallel to the surface \cite{Atwater,Pillai,Mallick,Mupparapu:15}. Also important results in absorption enhancement by random nanostructure can be framed in this aim by means of light diffraction \cite{holes,Pratesi:13}. It is interesting to note that the maximum value for mono-directional illumination is very close to the case of isotropic illumination: this result is indeed a further indication of the importance of the conversion to propagation along long dimensions (parallel to surface). The condition is reached by forcing the light in such direction by a proper value of scattering in the case of mono-directional illumination while in the isotropic case the condition is near intrinsically fulfilled. 
 
 \begin{figure}
        \centering\includegraphics[width=\columnwidth]{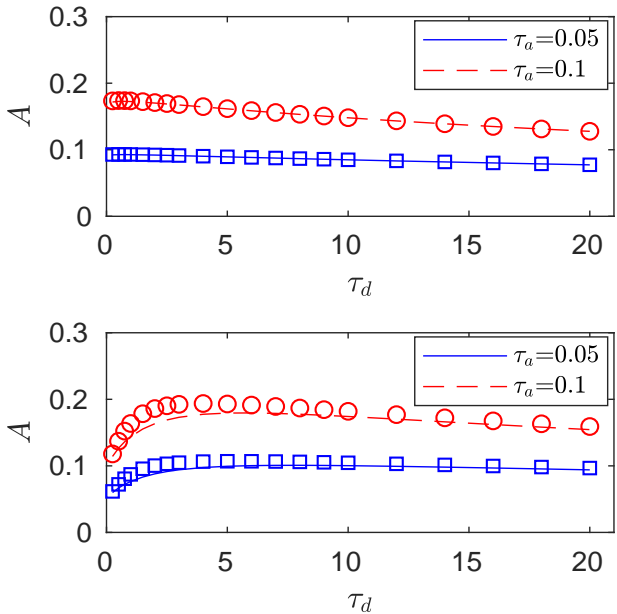}
        \caption{Absorption for different values of $\tau_d$ in a slab without refractive index mismatch with lambertian (top figure) and mono-directional (bottom figure) illumination. The lines are the case of $g=0$ and the markers are the one of $g=0.9$. A different g does not appreciably change the results.   }
        \label{fig:LT}
   \end{figure}

In Figs.~\ref{fig:LT} the absorption is reported in the case of two different values of the asymmetry parameter $g$ (0 and 0.9). The same behaviour for the two cases testifies that the reduced scattering coefficient $\mu_s'=\mu_s\,(1-g)=\tau_d/d$ is the fundamental parameter for the description of the scattering.

\subsection{Slab geometry with reflection}
\begin{figure}
        \centering\includegraphics[width=\columnwidth]{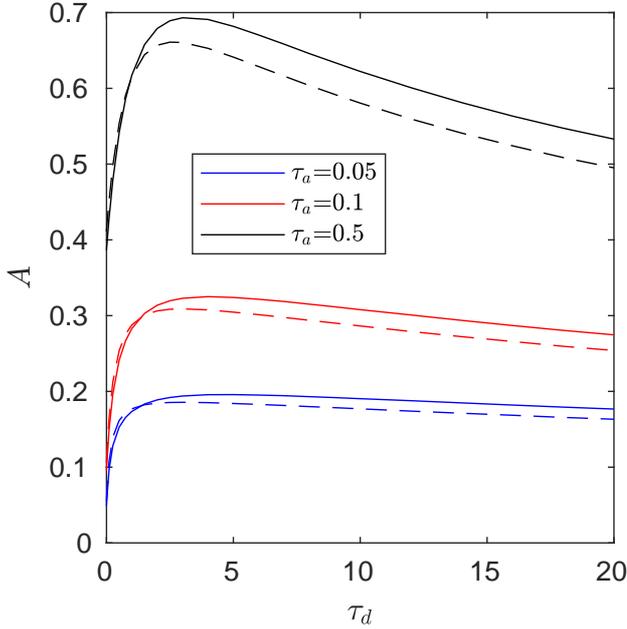}
        \caption{Absorption as a function of $\tau_d$ in the case of slab with refractive index mismatch ($n_r=n_2/n_1=1.525$) for different values of $\mu_a$. The continuous line is with  mono-directional illumination, while the dashed one is with lambertian illumination.}
        \label{fig:slab_R}
   \end{figure}

\begin{figure}
        \centering\includegraphics[width=\columnwidth]{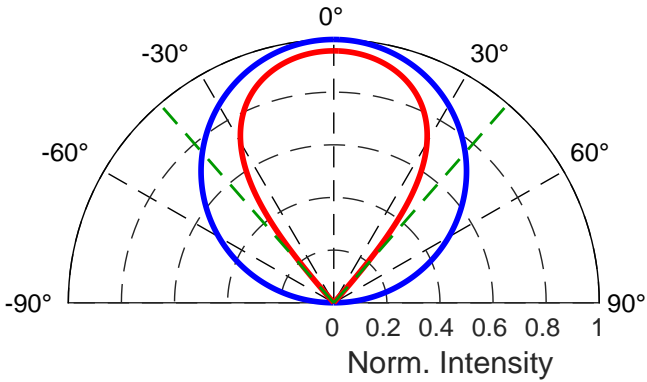}
        \caption{ Intensity pattern with a lambertian incidence on the interface. The blue  line is  the case of no refractive index mismatch, while the red continuous line is the case with $n_r=n_2/n_1=1.525$. The emission is normalized to the maximum of the former case. The dotted  green line is the limiting angle of emission in the last case. }
        \label{fig:lobo}
   \end{figure}

A further step to the enhancement of absorption by increasing the path inside the medium, it is the arrangement of the medium, within a material with reflecting surfaces due to the index of refraction mismatch between the internal  and the external media. As shown in Ref.~\cite{Savo765}, also in this case an invariance property holds: 
\begin{equation}
\langle L\rangle_{IP}=4\,{\left(\frac{n_{2}}{n_{1}}\right)}^2\frac{V}{S},
\label{inv_nr}
\end{equation}
where $n_2$ ($n_1$) is the index of refraction of the external (internal) material.
The mean length is then larger, respect to the slab without reflection, if $n_2>n_1$ as usually happens in a real device.
Also in this case one can wonder if the scattering can improve the absorption in case of isotropic or mono-directional illumination of a slab and the answer is positive as it is evident in Fig.~\ref{fig:slab_R}.  It is interesting to note that the two behaviours are similar, in contrast to the result for the slab without reflection. The explanation is in the actual similarity of the two kinds of illumination due to refraction at the interface. As shown in  Fig.~ \ref{fig:lobo} the lambertian lobe is indeed cut at large incidence angles because of the Fresnel reflection leading to a less ``isotropic'' and more directional-like illumination (besides the losses due to the reflection). As for all the other cases of illumination or considered geometry, the scattering is detrimental for absorption when it is very large. A proper choice of a moderate scattering strength maximizes the absorption with an enhancement that can be remarkable. These results are in agreement with a previous  study with a  mono-directional illumination for the same geometry \cite{Mupparapu:15}.

\section{Conclusions}
Summarizing our study on the effect of scattering strength on absorption, we found a remarkable enhancement for geometries with non homogeneous dimension (slab geometry) and conditions of non isotropic illumination. This behaviour is indeed coherently framed  by the search for a longer path inside the medium within the constraint given by the invariance property. In the case of mono-directional illumination the enhancement may occur because the scattering converts light entering the materials in long paths inside the medium, on the other hand long paths are possible when the dimension in a direction is much larger than the others as in a slab. 
In conclusion it is possible to tailor the conditions of illumination and the geometry of a medium to maximize its light absorption by scattering.

\section*{Acknowledgements}
Acknowledgements are due to prof.~Giovanni Zaccanti for his help in performing MC simulations. 

\appendix

\section{Derivation of Eq.~\ref{formula}}{\label{appen}}
Once known the pdf $P(L)$ of the total path length of the radiation inside the medium, the fraction of the absorbed radiation can be written as:
\begin{equation}
A=1-\int\limits_0^\infty P(L)e^{-\mu_aL}dL\,.
\label{formula1}
\end{equation}
In the case of low absorption ($\mu_a\,L\ll 1$), the exponential term can be expanded at the second order and integrated by using the definition of mean value:
\begin{equation}
A\approx \mu_a\,\langle L\rangle-\frac{1}{2}\mu_a^2\,\langle L^2\rangle\,.
\label{formula2}
\end{equation}
By using the definition of variance, the \ref{formula2} becomes:
\begin{equation}
A\approx \mu_a\,\langle L\rangle-\frac{{\mu_a}^2}{2}\,\left( \langle L \rangle^2 + \mathrm{Var}(L) \right)\,.
\label{formula3}
\end{equation}

\section*{References}


\end{document}